\documentstyle[aps,pra,amsfonts]{revtex}
\tightenlines
\newcommand{\be}{\begin{equation}}
\newcommand{\ee}{\end{equation}}
\newcommand{\ben}{\begin{eqnarray}}
\newcommand{\een}{\end{eqnarray}}
\newcommand{\bF}{\begin{figure}}
\newcommand{\eF}{\end{figure}}
\title{EPR Type Nonlocality in Classical Electrodynamics!}
\author{Partha Ghose and M. K. Samal}
\address{S. N. Bose National Centre for Basic Sciences,
Block JD, Sector III, Salt Lake, Kolkata 700 098}
\begin{document}
\maketitle
\begin{abstract}
It is shown that classical electrodynamics in its alternative
Kemmer-Duffin-Petiau-Harish-Chandra formulation surprisingly
reveals a Hilbert space structure leading to the possibility of
entangled states of classical radiation, and this in turn implies
the violation of Einstein-Bell locality in spite of Lorentz
invariance.
\end{abstract}
\vskip 0.2in

 \vskip 0.2in
\section{Introduction}
Ever since Bell \cite{bell} enunciated his famous theorem showing
that all local-realistic theories are incompatible with quantum
mechanics, it has been universally believed that the violation of
Einstein-Bell locality resulting from entanglement is a purely
quantum phenomenon without any classical analog. Entanglement is a
consequence of the Hilbert space structure of quantum mechanics
for multi-particle systems. Any theory having a Hilbert space
structure will necessarily imply entanglement and therefore
violation of Einstein-Bell locality. It therefore follows from
Bell's theorem that no local-realistic theory can have a Hilbert
space structure. The question is whether all classical theories
are necessarily local-realistic. Surprisingly, the answer turns
out to be negative. We will show in this paper that of all
classical theories electrodynamics, the quintessence of Einstein's
special theory of relativity and the epitome of locality, has a
Hilbert space structure that has so far been overlooked.

The classical Maxwell equations of electrodynamics can be written
in the form \cite{Harish,Kemmer}
\be
\beta_{\mu} \partial^{\mu} \psi(x) - i\l_0^{-1} \gamma
\psi(x)=0\label{eq:1}\ee where $\psi^T(x) = (1/\sqrt{2})( -E_x,
-E_y, -E_z, H_x, H_y, H_z, -l_0^{-1} A_x, -l_0^{-1} A_y, -l_0^{-1}
A_z, l_0^{-1} A_0)$, $ x=(t, {\bf x})$, $\l_0$ is a fundamental
length that drops out of all physical results, and the $\beta$'s
are the irreducible $10\times10$ representations of the
Kemmer-Duffin-Petiau (KDP) algebra\cite{comment1} \ben \beta_{\mu}
\beta_{\nu} \beta_{\lambda} + \beta_{\lambda} \beta_{\nu}
\beta_{\mu} &=& \beta_{\mu} g_{\nu \lambda} + \beta_{\lambda}
g_{\nu \mu}\label{eq:kdp} \een The matrix $\gamma$ is defined by
the relations $\gamma^2 = \gamma$ and
\be
\gamma \beta_{\mu} + \beta_{\mu} \gamma = \beta_{\mu}\ee It is
diagonal with the first six diagonal elements unity and the last
four diagonal elements zero. It projects out the four potentials
from $\psi$. Equivalently, multiplying the Lorentz covariant
equation (\ref{eq:1}) by $\beta_0$ and $(1 - \beta_0^2)$
respectively, one can replace it by the equations \ben
\frac{\partial (\gamma \psi)}{\partial t} = - c \tilde{\beta}_i
\partial_i (\gamma \psi)\label{eq:2}\een and
\be\beta_i \beta_0^2 \partial_i \psi - i l_0^{-1} (1 - \beta_0^2)
\gamma \psi = 0\label{eq:3} \ee where $\tilde{\beta}_i = \beta_0
\beta_i - \beta_i \beta_0$. Equation (\ref{eq:3}) is a first class
equation of constraint. These equations imply the massless
second-order equation
\be
\Box(\gamma \psi) = 0 \label{eq:box} \ee Equation (\ref{eq:1}) is
a combined form of the equations \ben \rm{curl}\, {\bf H} &=&
\frac{1}{c}\frac{\partial {\bf E}}{\partial t}\\ \rm{curl}\, {\bf
E} &=& - \frac{1}{c}\frac{\partial {\bf H}}{\partial t}\\ {\bf E}
&=& - {\bf \nabla} A_0 - \frac{1}{c}\frac{\partial  {\bf
A}}{\partial t}\\\partial^\mu A_\mu &=& 0\label{eq:mx1}\een
Equation (\ref{eq:2}) combines the first two Maxwell equations
above while equation (\ref{eq:3}) combines the constraint
equations \ben \rm{div}\, {\bf E} &=& 0\nonumber\\ {\bf H} &=&
\rm{curl}\, {\bf A}\label{eq:mx2} \een

We will henceforth use $\Psi \equiv \gamma \psi$ to denote the
classical wavefunction in electrodynamics. Since it is possible to
consider superpositions of $\Psi$ with complex coefficients, let
us define $\bar{\Psi} = \Psi^{\dagger} \eta$ where $\eta =
2\,\beta_0^2 - 1$, $\eta^2 = 1$ and $\eta \beta_0 = \beta_0$,
$\eta \beta_i + \beta_i \eta = 0$ \cite{Kemmer}. One can show that
the symmetric energy-momentum tensor can be defined as
\be
\Theta_{\mu \nu} = - \bar{\Psi} (\beta_\mu \beta_\nu + \beta_\nu
\beta_\mu - g_{\mu \nu})\Psi\ee so that
\be
- \Theta_{00} = \Psi^{\dagger} \Psi = \frac{\vec{E}.\vec{E} +
\vec{H}.\vec{H}}{2}\ee This is the physical significance of the
wavefunction $\Psi$: its norm is the energy density of the field.
One can also see that the local value of a physical observable $O$
can be defined by
\be
O =  \bar{\Psi} \hat{O} \Psi\label{eq:exp}\ee where $\hat{O}$ is
the corresponding operator. For example, the operators $c \tilde{
\beta}_i$ represent the components of the Poynting vector
$\vec{S}$:
\be
S_i = c \bar{\Psi} \tilde{\beta}_i \Psi = c (\vec{E} \times
\vec{H})_i\ee

It might appear so far that we have merely recast the familiar set
of Maxwell equations into an equivalent form, and that this is
devoid of any new physical implication. Surprisingly, this is not
the case. This formulation which is physically equivalent to the
standard Maxwell theory, helps reveal a hitherto unnoticed
underlying structure of classical electrodynamics, namely its
Hilbert space structure, and consequently the possibility of
entanglement and the violation of Einstein-Bell locality in
classical electrodynamics. This is what we will now demonstrate.

The first point to notice about this formalism is that it holds
only for massless fields. If one replaces $l_0^{- 1}$ in
(\ref{eq:1}) by $(m_0 c/\hbar)$, it becomes intrinsically quantum
mechanical\cite{ghose,pgmks}. Nevertheless, like $l_0^{- 1}$ the
fundamental mass parameter $m_0$ also drops out of all physical
results because of the operator $\gamma$.

The second point is the similarity with the Dirac algebra. The KDP
algebra has $126$ independent elements, and there are only three
inequivalent irreducible representations \cite{Kemmer}: $10^2 +
5^2 + 1^2 = 126$. As we have seen, the $10$ dimensional
irreducible representation of the KDP algebra is the relevant one
for electrodynamics. It is quite straightforward to see that the
operator
\be
\sigma_{\mu \nu} = [\beta_{\mu}, \beta_{\nu}]\label{eq:sigma}\ee
is the generator of transformations induced on $\psi(t,{\bf x})$
by the Lorentz transformations \cite{Kemmer}. This result will be
of importance in what follows.

Before proceeding further, it will be useful to remind the reader
of certain definitions in differential geometry and topology. The
space of all possible tangents at a point $p$ with local
coordinates $x^{\mu}(\mu = 1,...,n)$ on a manifold $M$ is called
the tangent space $T_{p}(M)$. It is spanned by the vectors
$\partial_{\mu}\equiv\partial/\partial x^{\mu}$, and is locally
isomorphic to ${\Bbb R}^n$. The union of the tangent spaces over
all points of $M$ is called the tangent bundle $T(M)$ over $M$.
This bundle is non-trivial if it cannot be reduced to ${\Bbb R}^n$
globally. A vector $D = a^{\mu}\partial_{\mu}$ is a called a field
on the tangent bundle $T(M)$. The dual of $T_{p}(M)$, denoted by
$T^*_{p}(M)$, is spanned by the vectors $d x^{\mu}$. The union of
all $T^*_{p}(M)$ is called the cotangent bundle $T^*(M)$. A
prototype field on $T^*(M)$ is $\omega = b_{\mu} dx^{\mu}$ and is
called a one-form.

Now notice equation (\ref{eq:1}). It can be written in the form
\be
D(t,{\bf x}) \psi(t,{\bf x}) = i l_0^{-1} \gamma \psi(t,{\bf x})
\label{eq:4}\ee where $D(t,{\bf x}) \equiv
\beta^{\mu}\partial_{\mu}$ is clearly a KDP algebra valued vector
field on the product bundle ${\cal P} \equiv S({\cal M})\otimes
T(\cal{M})$ over the orientable Lorentz manifold $\cal{M}$. To
identify $S(\cal{M})$, consider the group $SPIN(1,3)$ whose
elements are $\rm{exp}(i \sigma_{\mu \nu}\theta^{\mu \nu})$ with
$\sigma_{\mu \nu}$ defined by eqn. (\ref{eq:sigma}). In general,
$SPIN(1,3)= SL(2,C)$, the universal covering group of the Lorentz
group $SO(1,3)$ \cite{naka}. In this case, however, since no
spinor is involved, there is no difference between $SL(2,C)$ and
$SO(1,3)$, and $S(\cal{M})$ is simply the $SO(1,3)$ bundle over
$\cal{M}$. Note that although $T(\cal{M})$ is a trivial bundle
because ${\cal M}$ is a flat manifold, $S(\cal{M})$ is not because
the group manifold of the little group $SO(3)$ of the Lorentz
group is doubly connected, and hence ${\cal P}$ is a non-trivial
bundle.

The ten-component vector $\psi(x)$ is an irreducible
representation of the KDP algebra, but it is clearly a reducible
representation of the spin group $SO(1,3)$. The projection $(1 -
\gamma ) \psi$ consists of the four-vector potential which belongs
to the fundamental representation of $SO(1,3)$. The projection
$\gamma \psi = \Psi$ consists of the antisymmetric second rank
tensor field $F_{\mu \nu}$ which can be written as a complex
three-vector $(\vec{E} + i \vec{H})$ belonging to the fundamental
representation of the group $SO(3,C)$ which is homeomorphic to
$SO(1,3)$ \cite{xx}. Consider a unitary transformation $U \Psi =
\hat{\Psi}$ such that $\hat{\Psi}^\dagger = (1/2)(-E_x - i H_x,
-E_y - i H_y, -E_z - i H_z, E_x - i H_x, E_y - i H_y, E_z - i
H_z,0,0,0,0)$. $SO(3,C)$ acts on $\hat{\Psi}$. Therefore the
associated bundle is ${\cal B} = {\cal P}\times {\Bbb C}^3 \times
{\Bbb R}^4$, and $\psi(x)$ is a section of this bundle.

With this background it is evident that the $10$-dimensional
irreducible representation of the KDP algebra constitutes a
Hilbert space, being a complete linear vector space with a norm.
We will now show how this feature, hidden in the conventional
formulation of classical electrodynamics, plays a key role in the
violation of Einstein-Bell locality. Once a Hilbert space
structure exists, one can consider tensor product spaces
${\cal{H}}_{p1} \otimes {\cal{H}}_{p 2}$ and obtain
nonfactorizable states in such a product space. An example is
\be
\Psi (1,2) = \frac{1}{\sqrt{2}}\big[\Psi(1)\Psi(2) + \Psi(2)
\Psi(1)\big] \label{eq:pol1} \ee where \ben \Psi(1)^T &=& (-
E_x,0,0,0,H_y,0,0,0,0,0)\nonumber\\\Psi(2)^T &=& (
0,-E_y,0,-H_x,0,0,0,0,0,0)\een represent two plane polarized
electromagnetic waves whose polarization states along the $x$ and
$y$ axes are entangled. For convenience of notation let us denote
these states of linear polarization along the $x$ and $y$ axes
cryptically by $\hat{x}$ and $\hat{y}$ respectively. Then we can
write (\ref{eq:pol1}) as
\be
\Psi (1,2) = \frac{1}{\sqrt{2}}\,[ \hat{x}_1 \hat{x}_2 + \hat{y}_1
\hat{y}_2]\label{entg} \ee Such states will inevitably exhibit
EPR-Bell type nonlocal correlations and violate Bell's inequality
maximally. The `element of reality' of such a state is different
from the `element of reality' defined by EPR in terms of {\it
disturbance, prediction} and {\it probability} which are taken as
primary concepts in quantum mechanics. It is more akin to the
`element of reality' in the Bohmian sense in which these elements
are defined by the properties of the wave function and the set of
particles\cite{holland}. In the present case the `elements of
reality' are defined by the properties of the classical wave
function $\Psi$ alone, treated as a primary concept.

Until recently there was no evidence that macroscopic or classical
objects could be in entangled states. But recently a number of
experiments have shown that entangled states of coherent states in
optics and of macroscopic atomic objects do exist, though
transitorily because of environment-induced decoherence
\cite{nature}. The prevalent wisdom is that only quantum theory is
able to account for such states. Our analysis, on the contrary,
shows that even classical electrodynamics is able to account for
them in principle, though their production may be non-trivial.

Consider a classical state like (\ref{entg}) of two light beams
travelling in opposite directions along the $z$ axis. Let crystal
detectors be placed along their paths such that their polarization
axes make arbitrary angles $\alpha$ and $\beta$ with the $x$ axis
such that $\alpha - \beta = \theta$. Define the projection
operator \cite{peres}
\be
\sigma_{\theta} = (\hat{x} \hat{x}^{\dagger} - \hat{y}
\hat{y}^{\dagger})\,\,\rm{cos}\, 2 \theta +
(\hat{x}\hat{y}^{\dagger} + \hat{y}\hat{x}^{\dagger})\,\,
\rm{sin}\, 2 \theta\label{eq:corr}\ee Then, according to
(\ref{eq:exp}) the polarization correlation is given by \be <
\sigma_0 \otimes \sigma_{\theta}> =  \rm{cos}\, 2
\theta\label{eq:corr1}\ee exactly as in quantum mechanics. Thus,
if one observer has the choice of angles $\alpha$ and $\gamma$ on
one side and the other distant observer the choice between $\beta$
and $\gamma$, Bell's inequality takes the form \cite{peres}
\be
\vert \rm{cos}\, 2 (\alpha - \beta) - \rm{cos}\, 2 (\alpha -
\gamma)\vert + \rm{cos}\, 2 (\beta - \gamma) \leq 1\ee which is
violated by the correlation (\ref{eq:corr1}) for relative angles
of $30^{0}$ exactly as in quantum mechanics.

Before concluding, we would like to emphasize a few important
points.

1. An important difference between Hilbert spaces encountered in
classical physics and those encountered in quantum mechanics is
that the inner products in the former do not require a
probabilistic interpretation whereas inner products in the latter
do. This is because of wave-particle duality embodied in state
reduction in quantum mechanics. The wavefunction in quantum
mechanics is an unobservable position probability amplitude that
collapses to a point when a particle is detected at that point in
space at a certain time, whereas $\Psi(t,{\bf x})$ describes an
observable electromagnetic field spread over space-time.
Therefore, EPR-Bell nonlocality will be inherent in every
classical wave theory (both massive and massless) that can be cast
in a linear form similar to the KDP equation.

2. Contrary to prevalent wisdom, we have shown that {\it
Einstein-Bell nonlocality is consistent with Lorentz invariance}
which guarantees signal locality (the absence of superluminal
propagation\cite{comment}).

3. Finally, let us consider the bundle ${\cal B}_{12}= {\cal B}_1
\otimes {\cal B}_2$ corresponding to an entangled state. The
bundle coordinates are $(s, x, c_1, c_2, c_3, r_1, r_2, r_3, r_4)$
where $s \in SO(3,C)$, and there is no nonlocality in this
bundle-theoretic description. However, the projections in ${\cal
B}_1$ and ${\cal B}_2$ are respectively $\Pi_1(x)=x_1$ and
$\Pi_2(x)=x_2$. This shows that EPR-Bell nonlocality defined in
terms of separability on the base space (when $x_1 -x_2$ is
space-like) arises only when the description is given in terms of
the base space alone. This is the origin of this kind of
nonlocality. A similar conclusion would hold in quantum mechanics
also as is clear from the analogous mathematical structure of the
Dirac equation for which the spin group $S({\cal M})$ is $SL(2,
C)$ and the bundle is $S({\cal M}) \otimes E$ where $E$ is the
associated vector bundle of the principal bundle $P({\cal M}, G)$
with group $G$ in an appropriate representation \cite{comment2}.

To summarize, we have shown that (a) classical electrodynamics has
a Hilbert space structure embodied in eqn. (\ref{eq:4}) that leads
to the possibility of entanglement and violation of Bell's
inequality, (b) this EPR-Bell nonlocality exists only at the level
of Minkowski space but not in a bundle theoretic description, and
(c) such nonlocality is consistent with Lorentz invariance.
Although it might be difficult in practice to produce
polarization-entangled states of classical light required for
checking the violation of Bell's inequality, our analysis shows
that this should be possible in principle.

We are grateful to Partha Guha and Biswajit Chakraborty for
helpful discussions on fibre bundles and the spin group. We also
wish to thank the Department of Science \& Technology, Govt. of
India for a research grant that enabled this work to be
undertaken.


\begin{thebibliography}{99}
\bibitem{comment} However, it is still possible to have
Lorentz invariant superluminal tunneling, as shown in Ref.8.
\end{thebibliography}
\end{document}